%% file: main.tex
\documentclass[runningheads]{llncs}
\usepackage{graphicx}
\usepackage{multirow}
\usepackage{rotating}
\usepackage[indentfirst=false,rightmargin=0pt,leftmargin=0pt]{quoting}
\input{aux-commands}

\begin{document}
\title{The Integrated List of Agile Practices - A Tertiary Study}
%
%
\author{Michael Neumann\inst{1}}
\authorrunning{M. Neumann}
%
\institute{Hochschule Hannover, Ricklinger Stadtweg 120, 30459 Hannover, Germany\\
	\email{michael.neumann@hs-hannover.de}
}
\maketitle              
\begin{abstract}
\textit{Context:} Companies adapt agile methods, practices or artifacts for their use in practice since more than two decades. This adaptions result in a wide variety of described agile practices. For instance, the Agile Alliance lists 75 different practices in its Agile Glossary. This situation may lead to misunderstandings, as agile practices with similar names can be interpreted and used differently. \textit{Objective:} This paper synthesize an integrated list of agile practices, both from primary and secondary sources. \textit{Method:} We performed a tertiary study to identify existing overviews and lists of agile practices in the literature. We identified 876 studies, of which 37 were included. \textit{Results:} The results of our paper show that certain agile practices are listed and used more often in existing studies. Our integrated list of agile practices comprises 38 entries structured in five categories. \textit{Contribution:} The high number of agile practices and thus, the wide variety increased steadily over the past decades due to the adaption of agile methods. Based on our findings, we present a comprehensive overview of agile practices. The research community benefits from our integrated list of agile practices as a potential basis for future research. Also, practitioners benefit from our findings, as the structured overview of agile practices provides the opportunity to select or adapt practices for their specific needs.

\keywords{Agile practices  \and agile methods \and agile software development \and tertiary study}
\end{abstract}

\section{Introduction}
\label{Sec1}
The use of agile methods in software development has grown steadily over the past two decades \cite{VersionOne.2021}. More and more companies, regardless of their size or industrial sector, are using agile approaches. As a consequence, agile approaches are used in diverse settings. It follows that the use of agile methods and practices deviates from one another, which leads to several adaptions \cite{Noll.2019}. 

Various authors describe that agile methods such as Scrum or extreme programming (XP) are usually not fully adapted and used in companies (e.g., \cite{Diebold.2014,Williams.2010}). This statement follows Ken Schwaber, co-author of the Scrum Guide \cite{Schwaber.2020}. He assumed that 75 \% of all companies do not use Scrum as described in the Scrum Guide, but in an adapted approach \cite{Salo.2008}. According to Abrahmsson \cite{Abrahamsson.2003} the adaptation of agile methods and practices is often argued with the complexity of the agile transition. Stray et al. point to organizational aspects when introducing and adapting agile roles, artifacts, and practices \cite{Stray.2020}. This results in a high number of agile practices with many variants used in practice and described in literature. The Scrum and XP guidelines~\cite{Beck.2000,Schwaber.2020} describe 12 different practices, each. Furthermore, the Agile Alliance lists 75 different practices in its Agile Glossary \cite{AgileAlliance.2015}. Due to the combined use of agile practices of different agile methods such as Scrum, and increasingly on Lean approaches like Kanban, steadily new variants of agile practices are developed and used. 

This situation leads to the challenge of getting an overview of the agile practices used in diverse settings. In the past, secondary studies such as systematic mapping studies and systematic literature reviews were carried out in order to ascertain the state of the current research regarding agile practices in different contexts. These contexts include, for example, the affiliation of agile practices to methods and processes \cite{Williams.2010}, the use in different project-related contexts \cite{Diebold.2014} or in global software development \cite{Jalali.2012,Jalali.2010}. Due to their different research contexts and focus, the listed agile practices in these studies differ from one another. However, we did not find an integrated list of agile practices, which aims to provide a comprehensive overview of well-known agile practices described in recent literature and/or used in practice. This leads us to our two research questions: 

\begin{quote}
	\textbf{RQ 1:} \textit{Which agile practices are described and/or listed in the literature?}\\
	\textbf{RQ 2:} \textit{How can we synthesize the listed agile practices related to their characteristics/purpose?}\\
\end{quote}

This paper is structured as follows: First, we describe the background and related work of the study in Section \ref{Sec2}. We explain the selected research approach in Section \ref{Sec3}. An overview of the agile practices found in the literature is given in Section \ref{Sec4}. We present our approach for synthesizing the extracted agile practices from the literature and as the result our integrated list of agile practices in Section \ref{Sec5}. Before the paper closes with a conclusion in Section \ref{Sec7}, we discuss the limitations of our study in Section \ref{Sec6}. 

\section{Background and Related Work}
\label{Sec2}
Today, agile methods are well-known approaches in software development \cite{VersionOne.2021}. The idea of iterative and incremental approaches goes back to the 1950s \cite{Larman.2003}. In the past decades, agile methods are often understood as a reaction on plan-driven approaches like the waterfall model. For instance, this is argued due to their aim of fast time response on changes during the project period and their iterative structure \cite{Williams.2010}. According to Abrahamsson \cite{Abrahamsson.2002} agile methods are incremental, adaptable and cooperative approaches. 

Another aspect concerning agile methods is the value-based work and the strong focus on social aspects like collaboration and interaction. The agile manifesto defines a set of four value-pairs and twelve principles \cite{Beck.2001}. In addition, further values and principles are defined in guidelines for agile methods like the Scrum Guide for Scrum \cite{Schwaber.2020}. Also, other elements of agile methods like artifacts, roles and practices are described in these guidelines. Agile methods like Scrum or XP were created with the purpose to provide specific approaches for an agile transition and usage in software development. We know from the literature \cite{Kuhrmann.2017} and practice \cite{VersionOne.2021} that the adaption of agile practices (e.g., the combination of several agile practices from different methods) is the normal case. 

In order to find the related work of our study, we searched for surveys and systematic mapping studies or literature reviews (SLR) dealing with agile practices and, if available, provide a list of agile practices.

Several authors deal with agile practices in different contexts. Diebold and Dahlem present a systematic map of agile practices in practice \cite{Diebold.2014}. The authors focus on empirical studies dealing with the use of agile practice in software development projects. Also they present an overall usage of agile practices in software development projects. The used list of agile practices consists of 18 entries. 

Jalali and Wohlin investigate the use of agile practices in the field of global software engineering \cite{Jalali.2012,Jalali.2010}. Their studies focus on the use of 26 different agile practices in the several distribution types. Also, Camara et al. dealing with a similar topic in their systematic literature review on agile global software development \cite{Camara.2020}. They identified 48 different agile practices in use in that context.

The authors \cite{Camara.2020,Diebold.2014,Jalali.2012,Jalali.2010} also found, that agile practice were adapted and thus, customized agile methods were applied. Other studies only addressed sub-problems. For instance, Albuquerque et al. deal with agile requirements engineering \cite{Albuquerque.2020}. The authors considered 14 different agile practices in their mapping study. Sandstø and Reme-Ness investigating in their systematic literature review the relation of agile practices and their impact on project success \cite{Sandsto.2021}. The authors identified 12 agile practices and describe their impact on specific conditions for project success, such as communication or motivation of the team members.

However, to the best of our knowledge we did not find any study aiming to synthesize the variety of agile practices and provide an integrated overview of agile practices. Thus, we decided to conduct a tertiary study, which takes the findings from the recent literature into account. We present our research approach in the next section.

\section{Research Method}

\label{Sec3}
According to Petersen et al. \cite{Petersen.2008} systematic mapping studies are used to ascertain the current state of research in a field of interest in Software Engineering. The motivation of this study is to provide an overview of agile practices used and described in the literature and, based on this, to create a synthesized list of agile practices. From our point of view, the approach of a systematic mapping study is suitable for this purpose. Nonetheless, we have also used methods of the SLR guidelines of Kitchenham and Charters for conducting systematic literature reviews \cite{Kitchenham.2007}. This combined approach (for conducting systematic literature reviews and systematic mapping studies) has already been chosen by several authors in the past (e.g., \cite{Diebold.2014,Koskinen.2019}). To increase the traceability and transparency of our systematic mapping study, this approach appears to be useful.

As recommended by Kitchenham and Charters \cite{Kitchenham.2007}, we developed and used a protocol to document our study. The protocol contains the relevant information of the study including the research goal and questions, search strategy, study selection procedure and data extraction. We describe our approach in the following subsections based on the protocol. 

\subsection{Search Strategy}
We selected Scopus for applying our literature search. We decided to use Scopus as the library lists various publishers (such as SpringerLink, ACM, or Wiley). Besides, other authors have used Scopus for conducting systematic literature reviews and systematic mapping studies (e.g., \cite{Aleksander.2021,Stray.2020}). 

In a first step for developing the search string, we derived keywords and grouped them to keyword groups from our first research questions. Next, we connected the keyword groups with a Boolean operator and defined specific keywords for the related keyword group: $<$Agile practice$>$ AND $<$Agile software development$>$

Using our initial search string, we carried out test runs in Scopus and Google Scholar. During the test runs, we skimmed the results (title, keywords, abstract) and optimized the search string based on the findings, for example, whether keywords were missing. 
After several iterations, we defined our final search string, which we used for the search in Scopus:\\ \textit{(("agile practic*") AND ("agile" OR "agile software development" OR "agile method" OR "agile methods" OR "agile methodologies" OR "agile methodology" OR "lean software development"))}

The final search run was performed in June 2021 with an activated year range filter set to "since 2010". We argue the choice of a selected time range filter as our study aims to create an integrated list of agile practices based on the recent literature, including the actual state of usage of agile practices. The result set contained 876 potentially relevant studies. We used the Scopus interface to export the meta data of the studies and imported them to our data extraction file, which we created with Microsoft Excel.

\subsection{Study Selection}
In order to be able to perform the study selection it is recommended by Kitchenham and Charters \cite{Kitchenham.2007} to define inclusion and exclusion criteria. We defined three inclusion and eight exclusion criteria (see Table~\ref{Table:Studyselection}). The inclusion criteria IC1 was implicitly obtained by the activated search filter on year range setting when conducting the search in Scopus. Besides the structural exclusion criteria EC1 to EC4, we defined five content related exclusion criteria (EC5 to EC9).

We used the structural exclusion criteria EC1 to EC4 for an initial selection of the primary studies. During this check we excluded nine studies: Five, because of gray literature (EC1) and four, because the studies were not written in English (EC3).

\begin{table}[h]
\centering
\begin{tabular}{l | p{9.5cm}}
\hline
\textbf{Category} & \textbf{Criterion} \\
\hline
\multirow{4}{*}{Inclusion}
 & \textbf{IC1:} Studies published between 2010 and 2021\\
 & \textbf{IC2:} Studies written in English\\
 & \textbf{IC3:} Studies published in the field of agile software development\\
\hline
\multirow{4}{*}{Exclusion}
  & \textbf{EC1:} Gray literature (e.g., technical or experience reports)\\
  & \textbf{EC2:} Contributions with less than three pages\\
  & \textbf{EC3:} Studies not written in English\\
  & \textbf{EC4:} Studies not peer-reviewed \\
  & \textbf{EC5:} Studies not dealing with a list of specific agile practices\\
  & \textbf{EC6:} Studies focus on educational contexts (e.g., agile methods in higher educational)\\
  & \textbf{EC7:} Studies dealing with agile methods without a connection to software development\\
  & \textbf{EC8:} Studies dealing with software development and related topics without a connection on agile software development and agile practices in particular\\
\end{tabular}
\caption{Study selection criteria}
\label{Table:Studyselection}
\end{table}

\begin{figure}[htbp]
    \centering
	\includegraphics[scale=0.60]{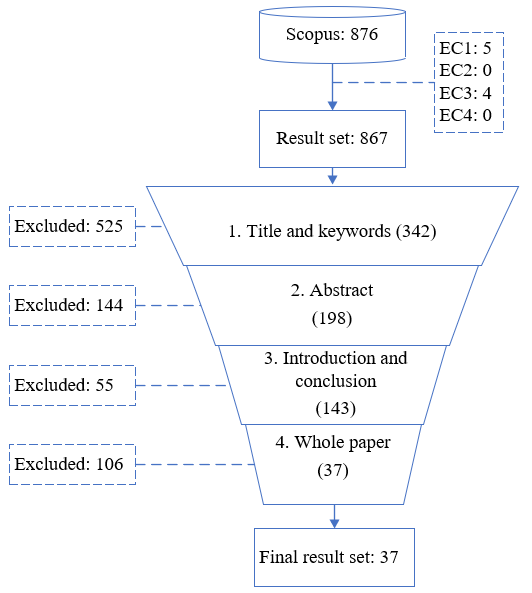}
	\caption{Results of the study selection process}
	\label{fig1}
\end{figure}

Based on the result set of 867 studies we performed a four stage study selection procedure\footnote{The protocol of our selection procedure is available at: https://sync.academiccloud.de/index.php/s/1nNipuDD655EJKF} (see Figure~\ref{fig1}). In the first step, we screened title and keywords of the respective study and excluded 525 studies. While reading the abstract in the second step, we excluded 144 studies. In the third step, reading the introduction and conclusion, we excluded 55 studies. During the fourth step, reading the whole content of the paper, we excluded 106 studies. The high number of removed studies in this step are due to the fact that any borderline cases left in the previous steps. The final result set contains 37 studies, which we used for data extraction. 

Most of the studies (757) were excluded because they are not dealing with a list of specific agile practices (EC5). Also, we excluded 49 studies, because they are focusing on educational contexts (EC6). For instance, the adaption of agile methods in higher education. Further 16 studies were excluded due to their missing connection to software development (EC7). Five studies were not dealing with agile methods in software development (EC8). Only three studies were duplicates.

\subsection{Data Extraction}
We read each paper of our result set of 37 studies completely in order to be able to extract the relevant information from the studies. We documented the data extraction in a Microsoft Excel file. The file contains general information like author/s or title of the study and specific data such as the research focus and method or the agile practices described in the study (see Table~\ref{Table2:DataExtraction}).
\begin{table}[h]
\centering
\begin{tabular}{p{3cm} | p{8cm}}
\hline
\textbf{Attribute} & \textbf{Information}\\
\hline
Author & General information\\
\hline
Title & General information\\
\hline
Year & General information\\
\hline
DOI & General information\\
\hline
Document Type & Conference paper or journal article\\
\hline
Research focus & Is the study focusing on practical aspects (like projects) or theoretical contributions (like descriptive models)\\
\hline
Research method & The research approach used in the paper (e.g., quantitative survey, case study, ...)\\
\hline
Agile Practices & The list of the agile practices described, named or used in the study.\\
\hline
\end{tabular}
\caption{Structure of the Data Extraction Sheet}
\label{Table2:DataExtraction}
\end{table}

The general information (author/s, title, year, DOI and document type) were extracted automatically based on the Scopus export file. The author checked the content of each attribute manually. In some cases the document type used to be corrected manually. The specific data (research focus, method and agile practices) were extracted manually for each paper. We extracted the agile practices in the form of lists, because in all studies several agile practices were named or used.

\section{Results of the Literature Review}

\label{Sec4}
\subsection{Overview of the Studies}
Before we discuss the results of the SLR and answer the research questions in the following subsection and section \ref{Sec5}, we give a structural overview of the studies.

\begin{figure}[tb]
\includegraphics[scale=0.60]{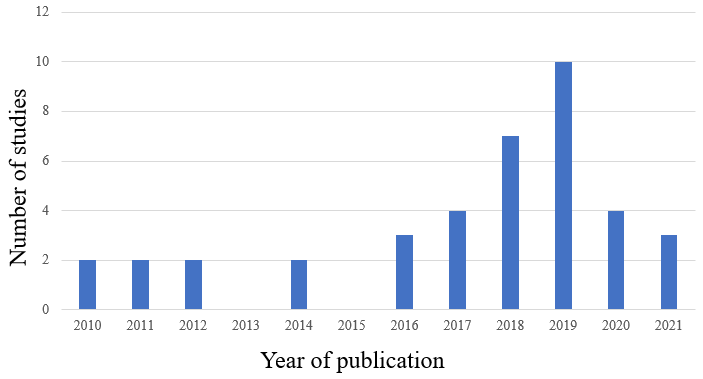}
\caption{Overview of the studies per publication year}
\label{fig2:StudiesPerYear}
\end{figure} 

The document type information had to be adjusted manually for the respective studies, as Scopus does not export this information correctly in some cases. The studies are published as conference papers (27) and articles in journals (12). Figure~\ref{fig2:StudiesPerYear} visualizes the distribution of the number of studies per year of publication. While only eight studies on this topic were published in the first five years of observation from 2010 to 2015, 29 studies have been published since 2016. Of these 29 studies, 24 studies have also been published since 2018, with only the first six months of 2021 being considered. Although a decrease can be determined in 2020 with only four publications, we have noticed an increased interest in the topic of agile practices.


There are several research methods used in the included studies. Eight studies use secondary research methods like systematic literature reviews (5) and systematic maps (3). The majority of the reported studies result from surveys (15). Also mixed approaches (8) and case studies (6) are often used by the authors.

\subsection{The current State of Agile Practices}
Based on the discussion in this subsection we answer our first research questions: \textit{Which agile practices are described and/or listed in the literature?}\\
First and foremost, our extracted data show a high variety and number of agile practices in use. In total, the 37 studies list 944 agile practices. The count of listed agile practices in the included studies range from 4 \cite{Pantiuchina.2017} to 93 agile practices \cite{Arcos.2020}. Almost half of the studies (17) list between 20 and 40 agile practices (see Figure \ref{fig3:DistributionofAP}). 

\begin{figure}[tb]
\includegraphics[scale=0.46]{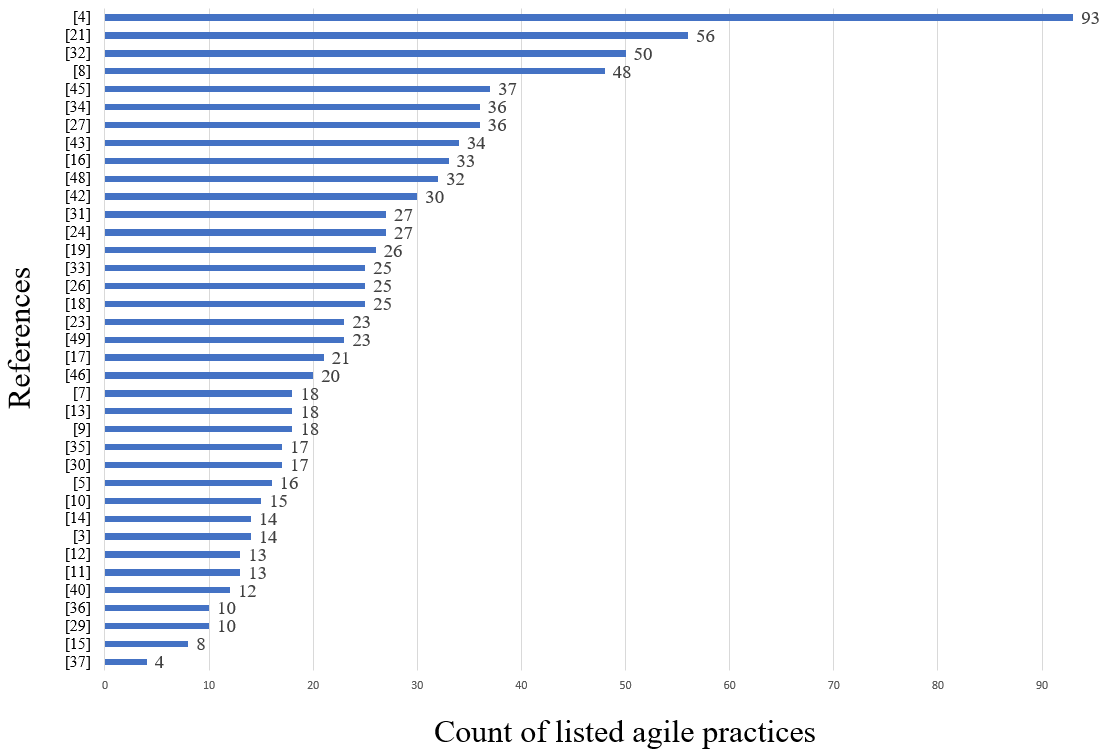}
\caption{Count of listed agile practices per study}
\label{fig:CountofListedAPperStudy}
\vspace*{-1em}
\end{figure} 

The agile practices listed in the studies are related to several characteristics. For example, various practices with a technical characteristic are used (such as refactoring or continuous integration). Also we found agile practices with an organizational characteristic like the office structure or energized work. However, it is not surprising that we found also collaborative focused practices such as daily stand up, planning, review or retrospective meetings. Interestingly, several studies describe/use/list these agile practices related to agile methods, especially Scrum and XP. 

The high variety of agile practices used in the literature is related to the research method and focus of the respective studies. The majority of the studies point to practical phenomena under study. Only one paper describes an overview of agile practices and methods \cite{Williams.2010}. We also analyzed the research focus. Here we found, that most of the studies (30) dealing with the usage of agile practices. Four studies each deal with the topics of adapting and adopting agile practices.

Although we identified a high variety of agile practices, we found several redundancies (same agile practice listed at least two times in different studies) of the listed agile practices in the included studies. Also, we identified that similar agile practices are listed, described or used under different names. Our handling with the redundancies in order to create a synthesized list of agile practices is described in the next Section~\ref{Sec5}.

\section{The integrated List of Agile Practices}
\label{Sec5}
\subsection{Synthesizing Agile Practices}
We answering our second research question in this subsection: \textit{How can we synthesize the listed agile practices related to their characteristics/purpose?}\\

We used the extracted data from the 37 studies as the basis for the procedure of synthesizing the lists of agile practices. For the synthesis, we created a new Microsoft Excel sheet and listed the agile practices of the 37 studies per column. We have also transferred the extracted information from the respective studies such as the title, author/s and year to the new Microsoft Excel sheet to ensure that the relevant information from the respective study to the list of agile practice is documented \footnote{The protocol of our synthesizing procedure is available at: https://sync.academiccloud.de/index.php/s/0YpKzzP56QBgmxU}. As mentioned in Section \ref{Sec4}, we identified various redundancies and found that the level of detail of the listed practices is different. This situation leads us to the following procedure: 

\paragraph{First step: Identify and remove the redundancies}
We removed any agile practice redundancies, we could find. An agile practice was marked as redundant when it is listed in at least two different studies. We also removed agile practices, if they did differ in name, but had essentially the same meaning. An example for this is the \textit{Daily meeting}, which is named and described as \textit{Standup Meeting} \cite{Jalali.2012,Jalali.2010,Williams.2010}, \textit{Daily discussion} \cite{Diebold.2014} and \textit{Stand Up}~\cite{Kurapati.2012}. Based on our findings we identified that the practices in some studies are on a more detailed level (e.g. according to Arcos-Medina~\cite{Arcos.2020}). The list of agile practices without any redundancies is the basis for the following steps.

\paragraph{Second step: Synthesize agile practices on an abstract level}
We screened the result list from step one in order to analyze the differences concerning the level of detail of the agile practices. We found, that the level of detail of the listed agile practices is heterogeneous. As a result, we identified that the majority of  agile practices is from a more detailed level. Thus, we decided to cluster the agile practices possible to a more abstract. The decision to cluster agile practices on an abstract level of detail was made, when we identified specific practices with the same purpose. Also, we mapped agile techniques (such as estimation techniques) to the agile practice on a more abstract level. This led to a more homogeneous level of detail across all agile practices on our list and provides clarity. 

For instance, we mapped testing practices from a more detailed level described in various studies (e.g., Test driven development, acceptance tests, automated testing or unit testing from Jalai and Wohlin \cite{Jalali.2010}) to the agile practice \textit{Agile Testing} in our list. We give another example with the agile practice \textit{Planning Game}. Here, we mapped specific estimation techniques such as \textit{Planning Poker} (e.g., from Williams \cite{Williams.2010}) and practices listed as \textit{Planning Game} (e.g., from Caires \cite{Caires.2018}).

\begin{figure}[tb]
\centering
\includegraphics[scale=0.55]{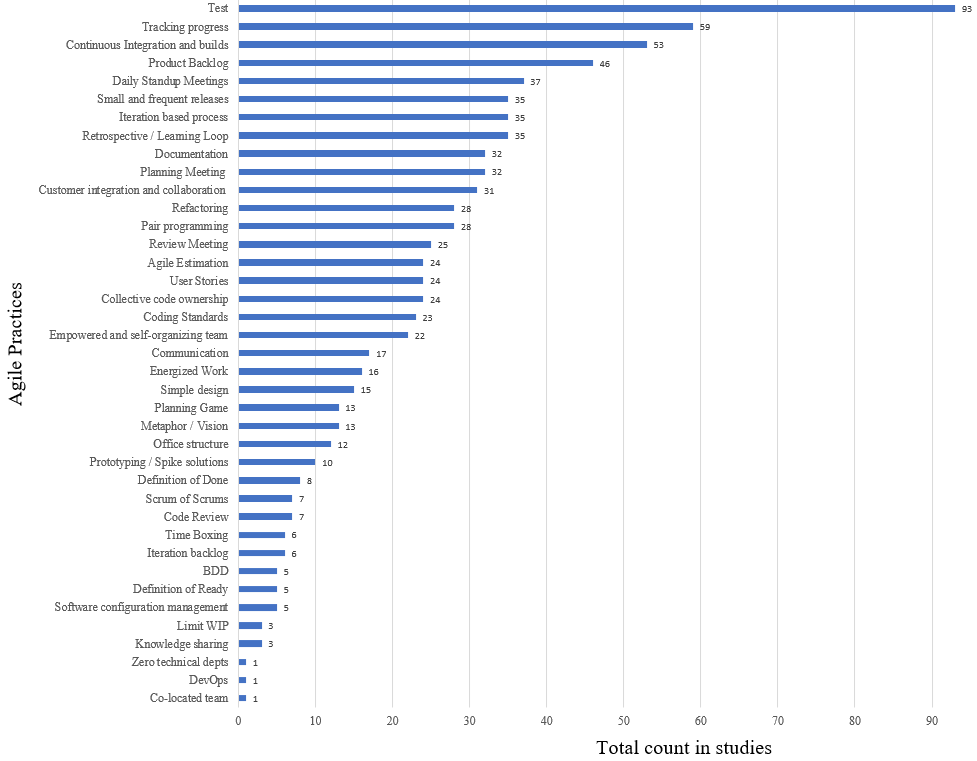}
\caption{Overview of the total count of mapped agile practice per synthesized practice}
\label{fig3:DistributionofAP}
\vspace*{-1em}
\end{figure}

Figure \ref{fig3:DistributionofAP} shows the number of the mapped (redundant, similar in terms of different names or level of detail) practices per synthesized agile practices in all included studies. 

The most mappings were conducted related to the agile practices \textit{Test}, \textit{Tracking progress} and \textit{Continuous integration and builds}. We identified more than 50 redundant or similar listed practices of these three agile practices, each. The high count of mapped practices to the synthesized agile practice Test is due to the several testing practices (e.g., Acceptance Test, Test Driven Development and Unit Testing), methods and approaches listed in the included studies.

\paragraph{Third step: Managing borderline cases}
We identified borderline cases during the step-by-step check of the redundancies (see step one) and the synthesizing on an abstract level (see step two). Some studies have listed practices that we did not classify as agile practices. For instance, Küpper et al. \cite{Kuepper.2019} list roles of agile methods such as Scrum Master or Product Owner. Also, agile methods such as SAFe \cite{Kuhrmann.2019} or Kanban \cite{Kropp.2016} are listed in several studies. In addition, methods and practices such as coaching \cite{Camara.2020} are described in the studies, which have non-related characteristics. We have marked these practices as borderline cases and checked them individually in this third step. During this check, we identified and documented a mapping to a few practices in our list (e.g., co-located team). In most cases, however, we have not added the borderline cases to our list of agile practices and not assigned them to practices that have already been listed.

\subsection{Introducing the integrated List of Agile Practices}
Before we introduce the integrated list of agile practices, we describe its structure and explain how we categorized the synthesized agile practices. 

\begin{figure}[tb]
\centering
\includegraphics[scale=0.6]{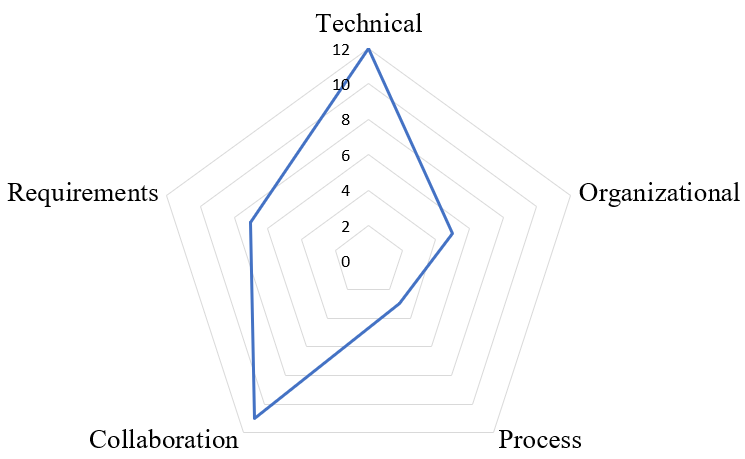}
\caption{Distribution of clustered agile practices per category}
\label{fig4:DistributionofAPperCat}
\vspace*{-1em}
\end{figure}

As explained in Sections \ref{Sec2} and \ref{Sec4}, the characteristics and purposes of agile practices differ from one another. In order to increase the clarity of our list of agile practices, we decided to categorize the agile practices. The categorization is based on the characteristics of the respective agile practices. In order to identify possible categories, we analyzed our list of agile practices entry per entry. We verified the agile practices characteristics mainly based on the guidelines from the well-known approaches Scrum \cite{Schwaber.2020} and XP \cite{Beck.2000}. Also, we used the glossary of agile practices from the Agile Alliance~\cite{AgileAlliance.2015}. However, some agile practices may relate to more than one category. This lay in the specific implementation of the respective agile practice. For example, a definition of done relates to a requirements characteristic, but also may be associated with a collaborative aspect as it is usually defined by the team. In order to follow our purpose to provide an integrate list of agile practices, we set the main characteristic described in the literature in focus. To minimize the risk of bias, we decided to conduct the categorization in three iterations. The first iteration of the categorization was conducted by the first author. In the second iteration two other researchers from the group did the categorization by themselves. In the following, we compared our results and discussed the very few mismatches we identified. In the final third iteration, we went through our categorization with four experts from the agile community and discussed the categorization for each agile practice. Below, we describe the five categories and provide examples. The distribution of the listed agile practices to the categories is presented in Figure~\ref{fig4:DistributionofAPperCat}.

As agile methods focusing on social facets like communication, it is not surprising, that we found several agile practices related to the characteristic of \textbf{collaboration}. In this category we assigned all agile practices concerning this characteristic. For example, agile practices within the team is collaborate closely together like in retrospective meetings. Furthermore, we found agile practices, which supports the collaboration. An example for this is the co-located team practice. In total, we added eleven agile practices to this category. 

We also found several agile practices with \textbf{technical} characteristics in our list. As technical associated agile practices are described for XP \cite{Beck.2000}, we assumed to find those in the literature. Examples for agile practices mapped to this category are coding standards, continuous integration and collective code ownership. Totally, 12 agile practices were mapped to the technical category.

Some agile practices concerning to an \textbf{organizational} characteristic. In this category, we clustered team-oriented agile practices like self-organization as well as other types of practices such as the office structure. Five agile practices are added to this category. 

Another facet of characteristics is related to a more \textbf{processual} background. This characteristic comes with agile practices as iteration based process. We added three practices this category.

Finally, we created the \textbf{requirements} category. In this category, we clustered all agile practices, which are related to any kind of requirements facets. These are, for example, using and maintaining a backlog as well as more detailed practices like the definition of ready or user stories. We added six agile practices to this category.

\newpage

The result of the synthesizing and categorizing process is the integrated list of agile practices. The list comprises 38 agile practices structured in five categories. We present the integrated list of agile practices in Table~\ref{Table:SListoAP}.

\begin{table}[ht]
\centering
\begin{tabular}{l|p{11.5cm}}
\hline
\textbf{Category} & \textbf{Agile practice (References)} \\
\hline
\multirow{12}{*}{Technical}
 & Agile Testing \cite{Arcos.2020,Bastarrica.2018,Caires.2018,Camara.2020,Diebold.2014,Diel.2017c2,Gren.2018,Heredia.2014,Jain.2016,Jalali.2012,Jalali.2010,Klotins.2021,Kropp.2016,Kuhrmann.2019,Kurapati.2012,Kuepper.2019,Lautert.2019,Mamoghli.2019,Myklebust.2018,Neto.2019,Noll.2019,Nurdiani.2019,Paez.2018,Pantiuchina.2017,Sletholt.2011,Souza.2019,Sanchez.2020,Tolfo.2018,Williams.2010,Yang.2019} \\
 & Code review \cite{Arcos.2020,Camara.2020,Kuhrmann.2019,Kuepper.2019,Noll.2019,Sanchez.2020}\\
 & Coding standards \cite{Arcos.2020,Bastarrica.2018,Caires.2018,Camara.2020,Heredia.2014,Jain.2016,Jalali.2012,Jalali.2010,Kropp.2016,Kuhrmann.2019,Kurapati.2012,Kuepper.2019,Lautert.2019,Licorish.2016,Mamoghli.2019,Myklebust.2018,Neto.2019,Noll.2019,Nurdiani.2019,Sletholt.2011,Sanchez.2020,Tolfo.2018}\\
 & Collective code ownership \cite{Arcos.2020,Bastarrica.2018,Caires.2018,Camara.2020,Diel.2017c2,Jain.2016,Klotins.2021,Kropp.2016,Kuhrmann.2019,Kurapati.2012,Licorish.2016,Mamoghli.2019,Myklebust.2018,Neto.2019,Noll.2019,Nurdiani.2019,Paez.2018,Sletholt.2011,Souza.2019,Sanchez.2020,Williams.2010}\\
 & Continuous integration \cite{Arcos.2020,Bastarrica.2018,Caires.2018,Camara.2020,Diebold.2014,Diebold.2017a,Diel.2017c2,Gabriel.2021,Gren.2018,Heredia.2014,Jain.2016,Jalali.2012,Jalali.2010,Klotins.2021,Kropp.2016,Kuhrmann.2019,Kurapati.2012,Kuepper.2019,Lautert.2019,Licorish.2016,Mamoghli.2019,Myklebust.2018,Neto.2019,Noll.2019,Nurdiani.2019,Paez.2018,Sandsto.2021,Sletholt.2011,Souza.2019,Sanchez.2020,Tolfo.2018,Williams.2010,Yang.2019}\\
 & DevOps \cite{Kuhrmann.2019}\\
 & Prototyping and Spike Solutions \cite{Albuquerque.2020,Kuhrmann.2019,Kuepper.2019,Mamoghli.2019,Myklebust.2018,Noll.2019,Sletholt.2011,Souza.2019,Sanchez.2020,Yang.2019}\\
 & Refactoring \cite{Arcos.2020,Bastarrica.2018,Caires.2018,Camara.2020,Diebold.2014,Diel.2017c2,Heredia.2014,Jain.2016,Jalali.2012,Jalali.2010,Klotins.2021,Kropp.2016,Kurapati.2012,Kuepper.2019,Lautert.2019,Licorish.2016,Mamoghli.2019,Myklebust.2018,Neto.2019,Noll.2019,Nurdiani.2019,Pantiuchina.2017,Sletholt.2011,Souza.2019,Sanchez.2020,Tolfo.2018,Williams.2010,Yang.2019}\\
 & Simple design \cite{Arcos.2020,Bastarrica.2018,Caires.2018,Camara.2020,Heredia.2014,Jain.2016,Klotins.2021,Kuepper.2019,Licorish.2016,Mamoghli.2019,Myklebust.2018,Nurdiani.2019,Sletholt.2011,Souza.2019,Tolfo.2018}\\
 & Small and frequent releases \cite{Arcos.2020,Bastarrica.2018,Camara.2020,Diebold.2014,Diebold.2019,Diebold.2017c1,Diel.2017c2,Gabriel.2021,Heredia.2014,Klotins.2021,Kropp.2016,Kurapati.2012,Mamoghli.2019,Myklebust.2018,Neto.2019,Paez.2018,Sandsto.2021,Sletholt.2011,Souza.2019,Tolfo.2018,Williams.2010,Yang.2019}\\
 & Software configuration management \cite{Arcos.2020,Klotins.2021,Kurapati.2012,Myklebust.2018,Souza.2019} \\
 & Zero technical depts \cite{Arcos.2020}\\
\hline
\multirow{12}{*}{Collaboration}
  & Agile estimation \cite{Albuquerque.2020,Camara.2020,Diebold.2017a,Diebold.2017c1,Diel.2017c2,Heredia.2014,Jain.2016,Klotins.2021,Kuhrmann.2019,Lautert.2019,Licorish.2016,Mamoghli.2019,Myklebust.2018,Neto.2019,Noll.2019,Souza.2019,Sanchez.2020}\\
  & Customer integration \cite{Arcos.2020,Caires.2018,Camara.2020,Diebold.2014,Gren.2018,Jain.2016,Jalali.2012,Jalali.2010,Kropp.2016,Kuhrmann.2019,Kurapati.2012,Kuepper.2019,Licorish.2016,Mamoghli.2019,Myklebust.2018,Noll.2019,Nurdiani.2019,Sandsto.2021,Sletholt.2011,Sanchez.2020,Tolfo.2018,Yang.2019}\\
  & Co-located team \cite{Kuepper.2019}\\
  & Communication \cite{Albuquerque.2020,Arcos.2020,Camara.2020,Diebold.2014,Jalali.2010,Kurapati.2012,Mamoghli.2019,Myklebust.2018,Nurdiani.2019,Tolfo.2018,Yang.2019}\\
  & Daily Standup Meetings \cite{Arcos.2020,Bastarrica.2018,Caires.2018,Camara.2020,Diebold.2014,Diebold.2017a,Diebold.2019,Diebold.2017c1,Diel.2017c2,Gabriel.2021,Gren.2018,Heredia.2014,Jain.2016,Jalali.2012,Jalali.2010,Klotins.2021,Kropp.2016,Kuhrmann.2019,Kurapati.2012,Kuepper.2019,Licorish.2016,Mamoghli.2019,Myklebust.2018,Neto.2019,Noll.2019,Nurdiani.2019,Pantiuchina.2017,Sandsto.2021,Sletholt.2011,Souza.2019,Sanchez.2020,Tolfo.2018,Williams.2010,Yang.2019}\\
  & Pair programming \cite{Arcos.2020,Bastarrica.2018,Caires.2018,Camara.2020,Diel.2017c2,Heredia.2014,Jain.2016,Jalali.2012,Jalali.2010,Klotins.2021,Kropp.2016,Kuhrmann.2019,Kurapati.2012,Kuepper.2019,Lautert.2019,Licorish.2016,Mamoghli.2019,Neto.2019,Noll.2019,Nurdiani.2019,Paez.2018,Sandsto.2021,Sletholt.2011,Souza.2019,Sanchez.2020,Tolfo.2018,Williams.2010,Yang.2019}\\
  & Planning Game \cite{Bastarrica.2018,Caires.2018,Camara.2020,Jalali.2012,Jalali.2010,Klotins.2021,Kurapati.2012,Neto.2019,Nurdiani.2019,Sletholt.2011,Williams.2010}\\
  & Release Planning \cite{Noll.2019,Sanchez.2020,Yang.2019}\\
  & Retrospective / Learning Loop \cite{Albuquerque.2020,Arcos.2020,Bastarrica.2018,Caires.2018,Camara.2020,Diebold.2014,Diebold.2017a,Diebold.2019,Diebold.2017c1,Diel.2017c2,Gabriel.2021,Gren.2018,Heredia.2014,Jain.2016,Jalali.2012,Jalali.2010,Klotins.2021,Kropp.2016,Kuhrmann.2019,Kurapati.2012,Kuepper.2019,Lautert.2019,Licorish.2016,Mamoghli.2019,Myklebust.2018,Neto.2019,Noll.2019,Nurdiani.2019,Paez.2018,Sandsto.2021,Sletholt.2011,Sanchez.2020,Tolfo.2018,Williams.2010}\\
  & Review Meeting \cite{Albuquerque.2020,Arcos.2020,Caires.2018,Camara.2020,Diebold.2014,Diebold.2017a,Diebold.2019,Diel.2017c2,Gabriel.2021,Heredia.2014,Jain.2016,Jalali.2012,Jalali.2010,Kurapati.2012,Kuepper.2019,Lautert.2019,Myklebust.2018,Neto.2019,Noll.2019,Sletholt.2011,Sanchez.2020,Tolfo.2018,Williams.2010}\\
  & Scrum of Scrums \cite{Camara.2020,Diebold.2017a,Diel.2017c2,Jain.2016,Klotins.2021,Noll.2019,Sanchez.2020} \\
\hline
\multirow{4}{*}{Process}
  & Iteration based process \cite{Arcos.2020,Caires.2018,Camara.2020,Diebold.2017a,Diebold.2017c1,Gren.2018,Heredia.2014,Jain.2016,Jalali.2012,Jalali.2010,Klotins.2021,Kuhrmann.2019,Kurapati.2012,Kuepper.2019,Licorish.2016,Mamoghli.2019,Myklebust.2018,Neto.2019,Nurdiani.2019,Paez.2018,Sandsto.2021,Sletholt.2011,Souza.2019,Tolfo.2018,Williams.2010,Yang.2019}\\
  & Limit WIP \cite{Heredia.2014,Noll.2019,Sanchez.2020} \\
  & Tracking progress \cite{Arcos.2020,Bastarrica.2018,Camara.2020,Diebold.2014,Diebold.2017a,Diebold.2017c1,Diel.2017c2,Gabriel.2021,Heredia.2014,Jain.2016,Jalali.2012,Jalali.2010,Klotins.2021,Kropp.2016,Kuhrmann.2019,Kurapati.2012,Kuepper.2019,Licorish.2016,Mamoghli.2019,Myklebust.2018,Neto.2019,Noll.2019,Nurdiani.2019,Sandsto.2021,Sletholt.2011,Souza.2019,Sanchez.2020,Tolfo.2018,Williams.2010}\\
\hline
\multirow{5}{*}{Requirements}
  & Behaviour Driven Development \cite{Diel.2017c2,Klotins.2021,Kropp.2016,Lautert.2019,Neto.2019,Souza.2019}\\
  & Definition of done \cite{Diebold.2019,Diebold.2017c1,Heredia.2014,Klotins.2021,Kuepper.2019,Myklebust.2018,Noll.2019,Sanchez.2020}\\
  & Definition of Ready \cite{Arcos.2020,Klotins.2021,Kuhrmann.2019,Kuepper.2019,Myklebust.2018,Sanchez.2020}\\
  & Documentation \cite{Camara.2020,Diebold.2017a,Gabriel.2021,Heredia.2014,Jalali.2012,Kurapati.2012,Kuepper.2019,Myklebust.2018,Noll.2019,Sandsto.2021,Souza.2019,Sanchez.2020,Yang.2019}\\
  & Metaphor / Vision \cite{Arcos.2020,Camara.2020,Diebold.2014,Gabriel.2021,Heredia.2014,Jain.2016,Jalali.2010,Kurapati.2012,Kuepper.2019,Nurdiani.2019,Sletholt.2011,Tolfo.2018}\\
  & User Stories \cite{Arcos.2020,Camara.2020,Diebold.2017a,Heredia.2014,Jain.2016,Jalali.2012,Jalali.2010,Klotins.2021,Kropp.2016,Kurapati.2012,Kuepper.2019,Myklebust.2018,Noll.2019,Sletholt.2011,Souza.2019,Sanchez.2020,Williams.2010,Yang.2019}\\
  & Using and maintaining a backlog \cite{Albuquerque.2020,Arcos.2020,Bastarrica.2018,Caires.2018,Camara.2020,Diebold.2014,Diebold.2019,Diebold.2017c1,Diel.2017c2,Gabriel.2021,Heredia.2014,Jain.2016,Jalali.2012,Jalali.2010,Klotins.2021,Kuepper.2019,Licorish.2016,Mamoghli.2019,Myklebust.2018,Neto.2019,Noll.2019,Sandsto.2021,Sletholt.2011,Souza.2019,Sanchez.2020,Tolfo.2018,Williams.2010,Yang.2019}\\
\hline
\multirow{5}{*}{Organizational}
  & Empowered and self-organizing team \cite{Arcos.2020,Camara.2020,Diebold.2014,Gabriel.2021,Heredia.2014,Klotins.2021,Kurapati.2012,Kuepper.2019,Neto.2019,Nurdiani.2019,Paez.2018,Sandsto.2021,Sletholt.2011,Souza.2019,Williams.2010,Yang.2019}\\
  & Energized work \cite{Arcos.2020,Caires.2018,Camara.2020,Heredia.2014,Jain.2016,Klotins.2021,Kurapati.2012,Kuepper.2019,Licorish.2016,Myklebust.2018,Sletholt.2011,Souza.2019,Tolfo.2018,Williams.2010,Yang.2019}\\
  & Knowledge sharing \cite{Diebold.2014,Kurapati.2012}\\
  & Office structure \cite{Caires.2018,Jalali.2012,Klotins.2021,Kropp.2016,Kurapati.2012,Licorish.2016,Myklebust.2018,Neto.2019,Sletholt.2011,Souza.2019,Tolfo.2018,Williams.2010}\\
  & Time Boxing \cite{Arcos.2020,Diebold.2014,Myklebust.2018,Nurdiani.2019,Souza.2019}\\
\hline
\end{tabular}
\caption{The integrated list of agile practices}
\label{Table:SListoAP}
\end{table}

\section{Limitations}
\label{Sec6}
Although we performed our study based on the guidelines by Petersen \cite{Petersen.2008} and according to Kitchenham and Charters\cite{Kitchenham.2007}, some limitations apply. A major challenge in systematic literature research is ensuring the completeness of the result set. To minimize the risk of omitting potentially relevant studies, we performed our test search runs in Scopus and Google Scholar. The search results showed high redundancies. Furthermore, several studies has proven the opportunity to work with Scopus as a single database for secondary studies (e.g., \cite{Aleksander.2021,Alsaqaf.2017,Stray.2020}). However, there is a possibility that we did not find all relevant studies due to the search being carried out in one database.

In addition, a limitation occurs due to the limited quality assurance of other researchers. The first author carried out the literature research, selection and data extraction by himself without systematic and iterative quality assurance measures by a second author. Therefore the potential risk arises that possible errors have been made while performing the literature search, e.g., optimizing the search string or selecting the studies due to bias. Similar limitations exist concerning the synthesizing procedure of the agile practices while creating the synthesized list of agile practices. We minimized these risks by performing cross-checks of our results by experts from the agile community and researchers from another research group. 

Another limitation relates to the selection of studies. We have defined various inclusion criteria, which have implicitly limited the result set. This concerns, for example, the limitation relating to the publication year. We have only considered results that were published since 2010. Even if our study shows that high redundancies in the naming of agile practices were already identified in the 37 included studies, it is conceivable that potentially relevant studies have already been published before. Furthermore, we have defined various exclusion criteria in order to be able to carry out and document the selection systematically and comprehensibly. It is also conceivable that we have excluded studies (e.g., due to non-English language) that are potentially relevant to the exclusion criteria.

\section{Conclusion and Future Work}
\label{Sec7}
This tertiary study was conducted with the purpose to create an integrated list of agile practices based on the literature to provide a comprehensive overview of agile practices. We analyzed 37 primary and secondary studies on detail in order to get an understanding of which agile practices are listed and/or used in the current state of research. We identified a high variety of agile practices related to the level of detail, which concerns due to the specific context of the respective studies. Furthermore, we found that various agile practices are listed redundantly in the included papers. 

In order to provide an integrated list of agile practices we decided to synthesize the agile practices extracted from the studies of our result set. The synthesize process consists of three steps. First, we removed the redundancies of the listed agile practices. The result of this first step was the basis for the upcoming procedure. Second, we analyzed the level of detail of the listed agile practices. We found that several agile practices are of a high level of detail while others are more abstract. To increase the clarity and simplicity we decided to cluster the agile practices to a more abstract level of detail wherever possible. Finally, we managed the borderline cases in the third step of our synthesizing procedure. 

After the synthesizing of the agile practices we structured our list of agile practices. The basis structure are five categories, which we identified by analyzing the 38 agile practices on detail. We implemented quality assurance measures for the categorization of the agile practices with the support of other researchers and practitioners from the agile community. The result of the two approaches is the integrated list of agile practices, which consists of 38 entries.

The findings of our study contributes to both, the research and practitioners community. For other researchers the integrated list of agile practices provides a comprehensive overview of agile practices used based on recent findings presented in the literature. The list of agile practices also contribute to a better understanding of the high variety of agile practices in practice, as almost of the included studies focus on practical phenomena under study. Thus, other researchers, which are dealing with agile practices may compare their findings with our integrated list of agile practices. 

We will use the integrated list of agile practices as a basis for our future work. In the next step, we aim to create a documentation of each agile practice. This documentation will provide more detailed information of the specific agile practices related to the purpose, their specific relation to agile method/s, a description and conceivable constraints. We also want to analyze to what extent the agile practices are related to one another to identify useful combinations of agile practices or even constraints. 




\bibliographystyle{splncs04}
\bibliography{references}

\end{document}

%% file: aux-commands.tex
\usepackage{ifthen}

\newboolean{auxinfo}
\setboolean{auxinfo}{true}   

\newboolean{showoutline}
\setboolean{showoutline}{true}

\newboolean{showcomment}

\setboolean{showcomment}{true} 

\usepackage[moderate]{savetrees}

\ifthenelse{\boolean{auxinfo}}
  {}
  {
    \setboolean{showoutline}{false}
    \setboolean{showcomment}{false}
  }

\usepackage{xcolor}
\usepackage{amssymb}

\ifthenelse{\boolean{showcomment}}
   {\newcommand{\nb}[2]{\fcolorbox{gray}{yellow}{\bfseries\sffamily\scriptsize#1}{\sf\small$\blacktriangleright${\em #2}$\blacktriangleleft$}}
   \newcommand{\working}[1]{\fcolorbox{gray}{yellow}{{\bf #1}\emph{\scriptsize---in progress---}}}
   \newcommand{\TBD}[1]{\fcolorbox{gray}{yellow}{{\bf #1}\textbf{TBD}}} 
  }
  {\newcommand{\nb}[2]{}{}
   \newcommand{\working}[1]{}
   \newcommand{\TBD}[1]{} 
  }

\usepackage{todonotes}
\ifthenelse{\boolean{showoutline}}{
	\newcommand{\outline}[3]{
		~\newline 
		\fcolorbox{red}{white}{
			\parbox{\columnwidth}{
				\ifthenelse{\equal{#1}{}}{
					\ifthenelse{\equal{#2}{}}{
						\noindent\colorbox[rgb]{0.65,0.16,0}{\textcolor[rgb]{1,1,1}{\textbf{Outline}}}
					}{
						\colorbox[rgb]{0.65,0.16,0}{\textcolor[rgb]{1,1,1}{\textbf{Outline -- Responsible: #2}}}
					}
				}{
					\ifthenelse{\equal{#2}{}}{
						\noindent\colorbox[rgb]{0.65,0.16,0}{\textcolor[rgb]{1,1,1}{\textbf{#1 page(s)}}}
					}{
						\colorbox[rgb]{0.65,0.16,0}{\textcolor[rgb]{1,1,1}{\textbf{#1 page(s) -- Responsible: #2}}}
					}
				}
				#3
			}
		}
	}
}{
	\newcommand{\outline}[3]{}
}

\newcommand\defauxcomm[1]{
       \expandafter\newcommand\csname #1\endcsname[1]{\nb{#1}{##1}}
       \expandafter\newcommand\csname WK#1\endcsname{\working{#1}}
       \expandafter\newcommand\csname TBD#1\endcsname{\nb{#1}}
    } 
   
\defauxcomm{KS}
\defauxcomm{MN}

\usepackage[normalem]{ulem}
\ifthenelse{\boolean{showcomment}}
    {\newcommand{\strike}[1]{\textcolor{red}{\sout{#1}}}}
    {\newcommand{\strike}[1]{}}